\documentclass[
	reprint,
	superscriptaddress,
	amsmath,amssymb,
	aps,
	pra,
	]{revtex4-1}

\usepackage{graphicx} % Include figure files
\usepackage{dcolumn} % Align table columns on decimal point
\usepackage[hidelinks]{hyperref} % add hypertext capabilities
\usepackage{soul}

\pdfoutput=1 % if your are submitting a pdflatex (i.e. if you have
\usepackage[T1]{fontenc} % if needed

\usepackage{benmacros} % see benmacros.sty
\usepackage[dvipsnames]{xcolor} % more colors

\hypersetup{pdftitle={To the Entanglement Wedge, and Beyond!}, pdfauthor={Ning Bao, Aidan Chatwin-Davies, Benjamin E. Niehoff, Mykhaylo Usatyuk}}

\newcommand{\mrm}[1]{\mathrm{#1}}

\newcommand{\Eq}[1]{Eq.~(\ref{#1})}

\newcommand{\Sec}[1]{Sec.~\ref{#1}}

\newcommand{\Fig}[1]{Fig.~\ref{#1}}

\newcommand{\Ref}[1]{Ref.~\cite{#1}}
\newcommand{\Refs}[1]{Refs.~\cite{#1}}
\newtheorem{thm}{Theorem}[section]

\newtheorem{defn}[thm]{Definition}

\begin{document}

\title{\boldmath Bulk Reconstruction Beyond the Entanglement Wedge}

\author{Ning Bao}
\email{ningbao75@gmail.com}
\affiliation{Center for Theoretical Physics and Department of Physics \\
	University of California, Berkeley, CA 94720, USA}
\affiliation{Computational Science Initiative\\Brookhaven National Lab, Upton, NY 11973, USA}
\author{Aidan Chatwin-Davies}
\email{aidan.chatwindavies@kuleuven.be}
\author{Benjamin E. Niehoff}
\email{ben.niehoff@kuleuven.be}
\affiliation{KU Leuven, Institute for Theoretical Physics\\
	Celestijnenlaan 200D B-3001 Leuven, Belgium}
\author{Mykhaylo Usatyuk}
\email{musatyuk@berkeley.edu}
\affiliation{Center for Theoretical Physics and Department of Physics \\
	University of California, Berkeley, CA 94720, USA}

\begin{abstract}
We study the portion of an asymptotically Anti de Sitter geometry's bulk where the metric can be reconstructed, given the areas of minimal 2-surfaces anchored to a fixed boundary subregion.
We exhibit situations in which this region can reach parametrically far outside of the entanglement wedge.
If the setting is furthermore holographic, so that the bulk geometry is dual to a state in a conformal field theory (CFT), these minimal 2-surface areas can be deduced from the expectation values of operators localized within the boundary subregion.
This presents us with an alternative: Either the reduced CFT state encodes significant information about the bulk beyond the entanglement wedge, challenging conventional intuition about holographic subregion duality; or the reduced CFT state fails to contain information about operators whose expectation values give the areas of minimal 2-surfaces anchored within that subregion, challenging conventional intuition about the holographic dictionary.
\end{abstract}

\maketitle
\flushbottom

\section{Introduction}
\label{sec:intro}
Over the last decade, the introduction of quantum information theory into the Anti de Sitter/Conformal Field Theory (AdS/CFT) correspondence \cite{Maldacena:1997re,Witten:1998qj} has had a profound impact on the field.
The most basic result of this program is arguably the Ryu-Takayanagi formula \cite{Ryu:2006bv}, which says that the Von Neumann entropy of a reduced state on a subregion of the boundary CFT is proportional to the area of the smallest bulk extremal surface that is homologous to the subregion.
More recently, investigations of the bulk region in between this extremal surface and the boundary subregion, called the \emph{entanglement wedge}, have led to deep results on holographic quantum error correction \cite{Almheiri_2015, Pastawski_2015}, the ability to reconstruct bulk perturbations of the vacuum within the entanglement wedge from boundary data \cite{Dong_2016, Cotler_2019}, and the equivalence of bulk and boundary relative entropy for perturbatively close states \cite{Jafferis_2016}.
Altogether, this paints a picture of ``subregion duality,'' according to which information about the bulk gravitational state, reduced on the entanglement wedge, is encoded in the subtended boundary region's reduced CFT state, and vice-versa.
Earlier work offers hints, however, that the full story may be more subtle \cite{Czech_2012, akers2019large}.

One such hint comes from comparing boundary-anchored extremal surfaces of differing dimensions.
In $d+1$ space-time dimensions, the dimensionality, $k$, of a boundary-anchored extremal surface can range from 1 to $d-1$. 
For different $k$, it is well known that extremal surfaces whose anchors lie within a fixed boundary subregion probe different regions of the bulk \cite{Hubeny:2012ry}.

In this letter, we will show that there exist situations in which boundary-anchored 2-dimensional extremal surfaces probe arbitrarily far outside of the entanglement wedge.
We then make two observations.
First, the areas of these surfaces are accessible from data in the boundary subregion.
In Refs.~\cite{Maldacena_1998,Ooguri_1999,Rey_2001}, it was argued that the expectation values of smooth, non-self-intersecting Wilson lines in the boundary CFT are given by the areas of 2-dimensional extremal surfaces anchored to the boundary.
This conjecture has been tested in several cases, and has since become the \emph{de facto} standard for computing such Wilson lines in holography.
Second, a recent result shows that knowing the areas of these surfaces is sufficient to uniquely fix the space-time metric in the region that they probe \cite{bao2019towards}.
We therefore arrive at an alternative:
Either 1. a boundary subregion contains information about much more of the bulk than just the entanglement wedge, or 2. a boundary subregion \emph{fails} to contain information about operators whose expectation values give the areas of extremal 2-surfaces anchored within that subregion, for example, smooth, non-self-intersecting Wilson lines.

\section{Extent of bulk minimal surfaces}
\label{sec:geometry}

Consider an asymptotically Anti de Sitter (aAdS) spacetime, $\mathcal{M}$, with $d+1$ space-time dimensions, which we take to be static so that we may restrict to a spatial slice.
Let $B_k$ be some simply-connected, $k$-dimensional submanifold in the boundary of the slice, where $1 \leq k \leq d-1$, such that the boundary of $B_k$ is non-empty: $\partial B_k \neq \emptyset$.
(One can take, e.g., $B_k$ to be a $k$-ball whose boundary is a topological $k$-sphere.)
Given this submanifold, consider the set of $k$-dimensional surfaces that share a boundary with $B_k$, that can penetrate into the spatial bulk, that are homologous to $B_k$, and whose areas are stationary with respect to deformations.
Let $m(B_k)$ denote the surface whose area is the smallest, or one such surface if there are many with the same minimum area.
For concreteness, if $k=1$, then $m(B_1)$ is a boundary-anchored geodesic; or, if $k=d-1$ and the space-time has a holographic CFT dual, then $m(B_{d-1})$ is the Ryu-Takayanagi surface for the boundary subregion $A = B_{d-1}$ \cite{Ryu:2006bv}.

The question that we begin with is the following: how far into the bulk does $m(B_k)$ reach?  The reach, or extent, of a boundary-anchored minimal surface was studied by Hubeny in aAdS space-times having planar symmetry in \Ref{Hubeny:2012ry}.
Concretely, working in the Poincar\'e patch, these are metrics of the form
\begin{equation}
\dd s^2 = \frac{1}{\tilde{z}^2} \left( - f(\tilde z) \dd t^2 + \dd x_i \dd x^i + h(\tilde{z}) \dd \tilde{z}^2 \right) ,
\end{equation}
where $i = 1, \dots, d-1$, the AdS boundary is at $\tilde z = 0$, and $f(\tilde z), h(\tilde z) \rightarrow 1$ as $\tilde z \rightarrow 0$.
Here, we follow Hubeny and use coordinates that are better-adapted to finding the bulk reach of minimal surfaces instead of the usual Fefferman-Graham coordinates \cite{AST_1985__S131__95_0}.
The coordinate $\tilde z$ labels the hyperbolic bulk direction, and so its largest value on $m(B_k)$, which we denote by $\tilde z_*$, characterizes the bulk depth to which $m(B_k)$ reaches.

We will be interested specifically in two cases: either an infinite strip with $k = d-1$, or a round $k$-ball for any $1 \leq k \leq d-1$.  The infinite strip is the region bounded by two (spatial) co-dimension-1 planes; from the coordinates $x^i$, label one coordinate by $x$ and the remaining $d-2$ coordinates by $y^i$, where $i = 2, \ldots, d-1$.
Then, with a proper choice for the origin and orientation of the coordinate system, an infinite strip of width $L$ is the boundary region described by the ranges $x \in [-L/2, L/2]$ and $y^i \in \mathbb{R}$.
Similarly, a $k$-ball is given by $\sum_{i=1}^k (x^i)^2 \leq R$ for some radius $R$, with the remaining $x^i = 0$ for $k+1 \leq i \leq d-1$.

In the case of pure $\mathrm{AdS}_{d+1}$, for which $f(\tilde z) = h(\tilde z) = 1$, the deepest reach of the minimal surface anchored to an infinite strip is given by \cite{Hubeny:2012ry}
\begin{equation} \label{eq:strip}
\tilde z_*^{\mrm{strip}} = L \frac{(d-1)}{\sqrt{\pi}} \frac{\Gamma(\tfrac{2d-1}{2d-2})}{\Gamma(\tfrac{d}{2d-2})}.
\end{equation}
For a $k$-ball, the minimal surface is a spherical cap for all values of $k$ whose reach is given by \cite{Hubeny:2012ry}
\begin{equation} \label{eq:disk}
\tilde z_*^{\mrm{ball}} = R.
\end{equation}
Away from pure AdS, one can expect more general behavior; however, what is important are these scalings of $\tilde z_*$ with respect to $L$ and $R$, which hold whenever the bulk metric is sufficiently close to pure AdS.

Next, we consider the following question: given a $(d-1)$-dimensional boundary subregion $A$, what part of the bulk can we reach with $k$-dimensional minimal surfaces that are anchored to submanifolds $B_k$ in the interior of $A$, where $k \leq d-1$?
Let us call such bulk regions $k$-wedges:
\begin{defn}
Let $A$ be a simply-connected, $(d-1)$-dimensional subregion in the boundary of a slice of a static aAdS${}_{d+1}$ space-time.
The \emph{$k$-wedge} of $A$, denoted $W_k(A)$, is the set of all points that lie on a $k$-dimensional minimal surface $m(B_k)$ for at least one simply-connected submanifold $B_k \subseteq A$, where $1 \leq k \leq d-1$.
\end{defn}
Again for illustration, in the case of a holographic space-time, $W_{d-1}(A)$ coincides with the entanglement wedge of $A$ when the latter contains no entanglement shadows \cite{Czech_2012,Headrick:2014cta}.

Different $k$-wedges for the same fixed subregion $A$ differ from each other.
This is not very surprising, and indeed, it is a straightforward consequence of Hubeny's investigations.
It is perhaps a bit more surprising, however, that in certain situations, different $k$-wedges can differ by an arbitrarily large bulk region.

\begin{figure}[ht]
\centering
\includegraphics[width=\linewidth]{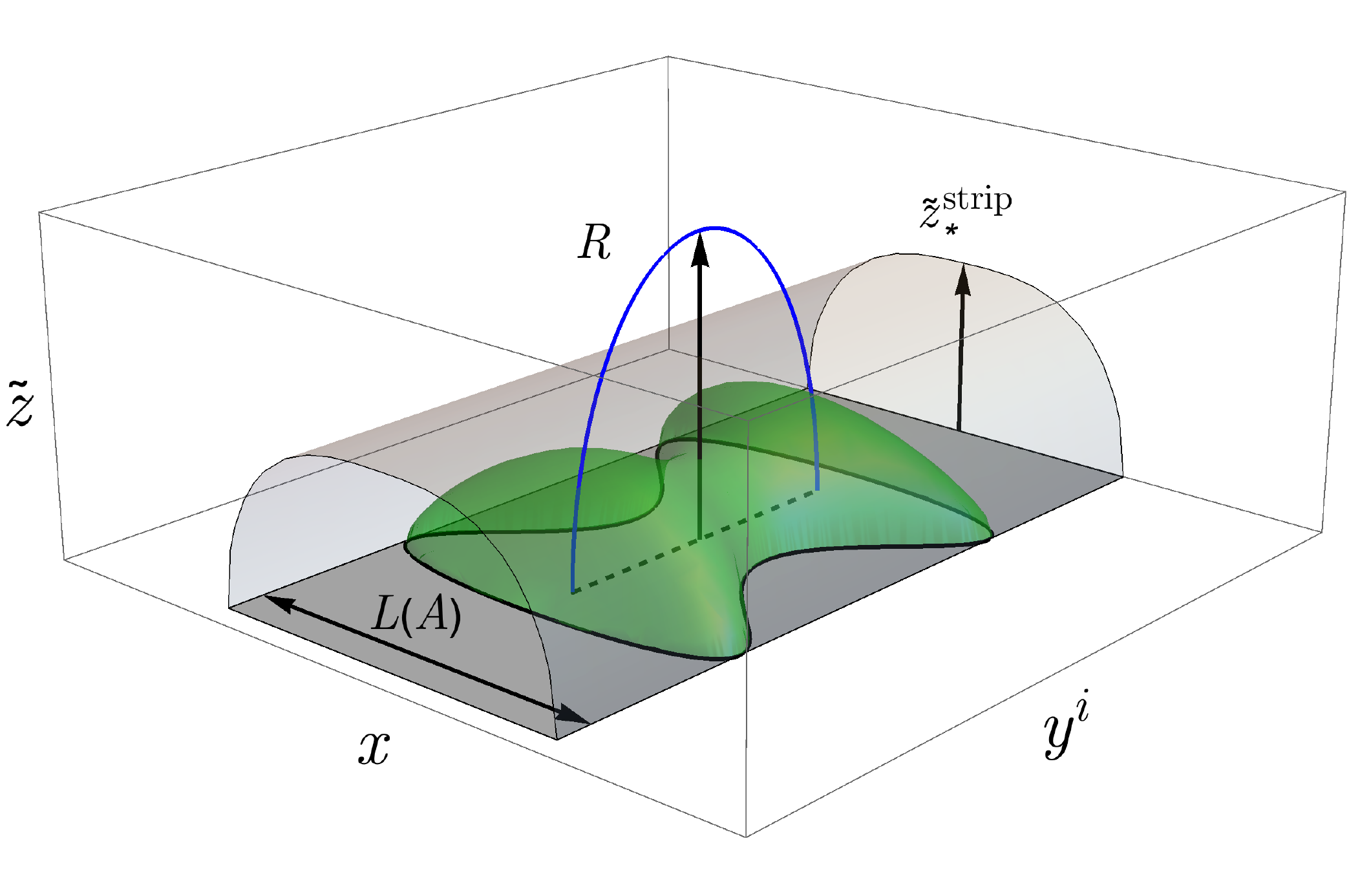}
\caption{Our construction, illustrated with $d-3$ dimensions suppressed. The boundary subregion $A$ can be inscribed in a $(d-1)$-dimensional strip of width $L(A)$. The minimal surface anchored to $A$ (drawn in green) is bounded by the minimal surface anchored to the strip (drawn in light grey), and so the bulk reach of $W_{d-1}(A)$ is less than $\tilde z_*^{\mrm{strip}}$. Nevertheless, there can be $k$-balls, such as the $k = d-2$ ball shown by the dashed line in this example, whose minimal surfaces extend well past $m(A)$ (or, equivalently, $\partial W_{d-1}(A)$) into the bulk.}
\label{fig:construction}
\end{figure}

We now construct one such situation, which we illustrate in \Fig{fig:construction}.
An important tool in our construction is the notion of the \emph{bounding width} of a subregion, given as follows:
\begin{defn}
	Let $A$ be a simply-connected, $(d-1)$-dimensional subregion in the boundary of a slice of a static aAdS${}_{d+1}$ space-time.
	The \emph{bounding width} of $A$, denoted $L(A)$, is the width $L$ of the smallest $(d-1)$-dimensional infinite strip which contains $A$.
\end{defn}
For example, if $d = 3$ and $A$ is the interior of an ellipse with semimajor axis $a$ and semiminor axis $b$, then the bounding width of $A$ is $L(A) = 2b$, the smaller of the two diameters.

Once again suppose that $\mathcal{M}$ is simply pure AdS${}_{d+1}$, and suppose that $d \geq 3$.
Let $A$ be a simply-connected, $(d-1)$-dimensional boundary subregion such that $\partial A \neq \emptyset$, and suppose that it has a bounding width $L(A) < \infty$.

Because a strip of width $L(A)$ contains $A$, it follows that the $(d-1)$-wedge of the strip contains the $(d-1)$-wedge of $A$.
This follows from entanglement wedge nesting \cite{Akers:2016ugt}, since in this case, the $(d-1)$-wedges of both $A$ and the strip coincide with these subregions' entanglement wedges.
It therefore follows that the deepest reach of $m(A)$ is bounded by \Eq{eq:strip}.

Now consider inscribing a round $k$-dimensional ball inside $A$, with $k \leq d-2$.
(A 1-dimensional ball is just a line segment.)
In particular, if $A$ is sufficiently larger than it is wide in at least $k$ directions, then we can inscribe a $k$-ball of radius $R > \tilde{z}_*^{\mrm{strip}}$.  Then according to \Eq{eq:disk}, the bulk reach of such a ball will exceed the bulk reach of the strip, and hence also that of $m(A)$.

However, nothing prevents us from choosing a boundary subregion that is arbitrarily longer than it is wide.
It will remain possible to inscribe such a region in an infinite strip of width $L$, and so the extent of its $(d-1)$-wedge will remain bounded by $\tilde z_*^{\mrm{strip}}$.
At the same time, based on being able to inscribe $k$-balls of arbitrarily large radius transversely to $A$'s bounding width, we can arrange for the subregion's $k$-wedges to probe arbitrarily deeply into the bulk.

To summarize, if a $(d-1)$-dimensional boundary subregion $A$ in AdS${}_{d+1}$ can be inscribed with lower-dimensional balls whose radii are sufficiently larger than the bounding width of $A$, then the $k$-wedges of $A$ will generically reach deeper into the bulk than $W_{d-1}(A)$ for $1 \leq k \leq d-2$.
Furthermore, we can immediately construct situations to which this observation applies in general static aAdS space-times as well.
Since $A$ can be chosen such that its bounding width is arbitrarily small, the minimal surface of the infinite strip which bounds $m(A)$ can be made arbitrarily close to the pure AdS case, due to the asymptotic AdS boundary conditions.
Therefore, the bulk depth to which $W_{d-1}(A)$ reaches remains bounded by $\tilde z_*^{\mrm{strip}}$ (plus subleading corrections that are calculated in \Ref{Hubeny:2012ry}), while $W_k(A)$ can be arranged to reach as deep into the bulk as one wishes, being obstructed only by topological barriers such as horizons.

\section{Bulk minimal surfaces and holographic reconstruction}
\label{sec:holography}

Now let us suppose that $\mathcal{M}$ is holographic, meaning that it is dual to a state in some $d$-dimensional CFT, which we can think of as living on $\partial \mathcal{M}$, in the large-$N$ and large-$\lambda$ limit. 
Now we can ask: what are the holographic consequences of the geometric observation that $k$-wedges can probe arbitrarily deeper than the entanglement wedge for $k \leq d-2$?

First, we note that the observation has an operational meaning, since data about certain minimal $k$-dimensional surfaces whose anchors lie in $A$ is accessible from the reduced CFT state $\rho_A$.
As an example, consider boundary Wilson loops contained in $A$.
Since $A$ is simply-connected, a Wilson loop can be specified as the boundary $\partial B$ of some disk $B \subseteq A$, and its expectation value is conjectured to be computed by a bulk path integral of the form \cite{Maldacena_1998, Ooguri_1999, Rey_2001}
\begin{equation} \label{eq:WilsonMaldacena}
\langle W(\partial B) \rangle = \int_{\sigma \sim B} \mathcal{D}\sigma ~ e^{-\sqrt{\lambda}S[\sigma]} ~ .
\end{equation}
The integral is over all 2-surfaces $\sigma$ that end on the Wilson loop.
In the limit where $\lambda = g^2 N$ is large, the above integral is approximated to leading order by its saddle point to obtain
\begin{equation} \label{eq:holoWilson}
\langle W(\partial B) \rangle =  \Delta \, e^{-\sqrt{\lambda} \, \mathcal{A}[m(B)]} ,
\end{equation}
where $\mathcal{A}[m(B)]$ is the regularized area of the minimal 2-surface anchored to $B$, and $\Delta$ is a prefactor whose contribution is subleading at large $\lambda$ (except in the presence of kinks or self-intersections, which will not appear in the theorem below \cite{bao2019towards}). 
Therefore, the areas of minimal 2-surfaces anchored within $A$ can be deduced from the reduced density matrix $\rho_A$ by extracting the expectation values of Wilson loops.
Similar expressions hold which relate boundary two-point functions to the length of boundary-anchored geodesics \cite{Louko_2000}.
In principle there may also be similar relations between higher-dimensional surface operators in the boundary and the areas of higher-dimensional boundary-anchored minimal surfaces; however, we are not aware of results that suit our purposes.

We focused on smooth boundary Wilson loops in particular because knowing the areas of all minimal 2-surfaces whose anchors lie on smooth, closed curves within $A$ is sufficient to uniquely fix the \emph{full bulk metric} in $W_2(A)$ for a large class of boundary subregions.
This is essentially the content of a recent theorem \cite{bao2019towards}:
Let $d \geq 3$, and let $A$ be a $(d-1)$-dimensional boundary subregion that is topologically a ball.
Suppose that every point in $W_2(A)$ lies on a minimal 2-surface $m(B)$, which can be retracted via a family of minimal surfaces to a single point in $A$.
In other words, for each $p \in W_2(A)$, suppose that there exists a continuous family of minimal 2-surfaces $\{m(S(\lambda))~|~\lambda \in [0,1]\}$, such that $S(\lambda) \subset \bar{A}$, $S(1) = B$, $p \in m(B)$, and $S(0)$ is a single point in the closure, $\bar{A}$ \footnote{In general, one must consider extremal 2-surfaces, but here we work with minimal 2-surfaces since we have restricted to a static spatial slice. The condition that every point in $W_2(A)$ be reachable by a continuous family of minimal surfaces which begins at a boundary point is necessary for the theorem. In particular, it rules out, e.g., 2-wedges that contain a compact horizon, where the minimal surfaces in any family which reaches a point on the other side of the horizon with respect to the boundary subregion will experience a discontinuous jump when the minimal surface wraps the horizon.}.
Then, knowing the areas of minimal 2-surfaces anchored to smooth, closed curves in $\bar{A}$ guarantees a unique reconstruction of the bulk metric in $W_2(A)$.

Therefore, via their entry in the holographic dictionary, knowing the expectation values of Wilson loops localized within $A$ ultimately ensures that it is possible to reconstruct the bulk metric in the region $W_2(A)$, provided that $W_2(A)$ satisfies the foliation condition in the theorem above and that it is possible to pass from \Eq{eq:WilsonMaldacena} to \Eq{eq:holoWilson} via a saddle point approximation.
When $d=3$, the bulk region foliated by 2-surfaces will in general be equal to or contained by the entanglement wedge.
When $d \geq 4$, however, according to our earlier geometric observation, we can generically find subregions $A$ such that $W_2(A)$ is both metric-reconstructible and extends parametrically far outside of the entanglement wedge.

\section{Discussion}
\label{sec:discussion}

The reasoning above leads us to two contrastive possibilities.  Either
\begin{enumerate}
\item There exists data in the reduced CFT state of a boundary subregion $A$ that can be used to reconstruct the bulk metric parametrically far outside the entanglement wedge of $A$, or
\item Such data (as would correspond to the areas of minimal 2-surfaces in a foliation beginning at a boundary point and ending outside the entanglement wedge) does not exist in the reduced CFT state.
\end{enumerate}
The first possibility conflicts with our intuition about subregion duality.
However, the second possibility conflicts with our understanding of the holographic dictionary entry for Wilson lines. 
Moreover, since minimal 2-surfaces that are anchored to a boundary subregion can be made to reach arbitrarily far outside of the entanglement wedge, the conflict cannot be resolved via some subleading quantum correction.
It is leading order geometric information.

Let us consider the first possibility.  According to the conventional lore, a boundary subregion is dual to its entanglement wedge, and vice-versa.  This can be precisely captured by, e.g., the formulation of Dong, Harlow, and Wall \cite{Dong_2016}.
According to this formulation, a boundary subregion and its entanglement wedge are dual in the sense that a sufficiently small algebra of bulk operators in an entanglement wedge, whose geometry is fixed to leading order in $N$ (the ``code subspace''), can be represented by a corresponding set of CFT operators in the boundary subregion.
Our argument, however, merely points to the possibility of reconstructing the bulk metric (and, it stands to reason, a bulk classical gravity background), and not the full physics.  Therefore it is not logically inconsistent with such precise formulations of bulk reconstruction, as the information recovered from the areas of minimal 2-surfaces need only relate expectation values of boundary operators to bulk quantities.
In particular, no operator relations are implied by our analysis.

The second possibility sounds rather attractive, given the precise formulations known which relate a boundary subregion $A$ to its entanglement wedge.  However, it leads to some surprising consequences.  One could conclude that the expectation values of Wilson lines simply do not correspond to the areas of minimal 2-surfaces.  While \Refs{Maldacena_1998, Ooguri_1999, Rey_2001} do find cases in which this correspondence fails, their findings only apply to pathological curves with kinks or self-intersections, whereas our argument here makes use of only smooth curves.
As such, one is forced to conclude that the duality between Wilson line expectation values and minimal 2-surface areas in the cases that we consider is merely an accident.
For instance, it could be that the duality breaks down in non-static settings in ways that are analogous to the breakdown of the geodesic approximation for Lorentzian correlation functions, as discussed in \Ref{Louko_2000}.
The question of whether such breakdowns in the saddle point approximation equating Wilson loop expectation values with areas of minimal 2-surfaces is an open one, which we hope to explore in future work.
Another option is that the duality could only hold in analytic space-times \cite{Giddings:2001pt,Freivogel:2002ex}.
Because the metric reconstruction theorem holds for non-analytic space-times, however, \Eq{eq:holoWilson} would have to break down at even the slightest non-analyticity.
Another, and perhaps stranger, possibility is that the presence or absence of Wilson line data in a reduced CFT state of a boundary region $A$ is dependent on the \emph{size} of the curve on which the Wilson line is defined, such that curves are excluded whenever a minimal 2-surface anchored on that curve would reach outside the entanglement wedge.
All of these limitations would also apply to any hypothetical object in the boundary subregion which encodes data about minimal 2-surface areas.

One question that our analysis immediately motivates is how to think of \emph{boundary} reconstruction.
Given access to only the entanglement wedge, what information is learned about the reduced CFT state?
Alternatively, what portion of the bulk is necessary to reconstruct the full reduced density matrix for a boundary subregion?
There have been hints in earlier works, such as \Refs{Evenbly_2017,Czech_2012}, that one can affect the reduced density matrix of the boundary subregion by acting outside of the entanglement wedge, and our result only reinforces these hints.
It would therefore be an interesting challenge to construct a systematic method of reconstructing the boundary reduced density matrix, in a sort of inverse problem to \Ref{Dong_2016}.

In this letter, we restricted ourselves to two-dimensional surfaces to exploit the metric reconstruction results of \Ref{bao2019towards}.
Metric reconstruction may also be possible using the area data of minimal $k$-surfaces, which would lead to similar conclusions for CFT operators whose expectation values correspond to such areas.  For $k=1$, these are geodesics, and the question of reconstructing the bulk metric from them is known in the mathematics community as the boundary rigidity problem \cite{Porrati:2003na,Croke2004,Pestov2005}.
While no such theorem is currently known for aAdS space-times, reconstruction using geodesic data deduced from boundary correlators would be an obvious extension of our work \footnote{In the context of geodesic probes, it would be important to ensure that one can construct a situation analogous to the one given here that does not run afoul of the limitations detailed in \Refs{Louko_2000,Engelhardt:2014mea}, according to which the geodesic approximation for boundary correlators breaks down. We thank Simon Ross and Sebastian Fischetti for stressing this to us.}.

Similarly, if the metric reconstruction theorem could be extended to disjoint boundary subregions, then 2-wedge reconstruction could potentially be thought of as a way of smoothing out the discontinuous jump that the entanglement wedge experiences across entanglement entropy phase transitions.
While the number of connected components in the entanglement wedge changes across such a transition, the number of connected components in the 2-wedge remains the same.
That the metric reconstruction theorem of \Ref{bao2019towards} is only proven for simply-connected boundary subregions is also our reason for focusing on these.

Another potential direction for further work is to study the problem in a time-dependent setting.
Some of the pieces are already in place, as the metric reconstruction theorem holds covariantly.
However, it would still be necessary to extend our construction in \Sec{sec:geometry}.
This may follow from careful application of the maximin proposal \cite{Wall_2014}, but the precise details remain to be seen.

\section{Conclusion}
\label{sec:conclusion}

In three or more bulk spatial dimensions, we observed that the bulk region swept out by spatial co-dimension 1 minimal surfaces anchored to a given boundary subregion can be significantly shallower than the regions swept out by the higher co-dimension minimal surfaces.
In four or more dimensions, holographically, this leads to an alternative.
Either one obtains knowledge about the bulk metric far outside of the entanglement wedge, given access only to the reduced CFT state on the wedge's boundary subregion; or the reduced CFT state somehow fails to contain complete holographic information, notably minimal 2-surface areas via the expectation values of Wilson lines in that subregion.

\acknowledgments
We would like to thank Nikolay Bobev, Raphael Bousso, Charles Cao, Venkatesa Chandrasekaran, Anthony Charles, Newton Cheng, Sebastian Fischetti, Temple He, Cindy Keeler, Alex Maloney, Jason Pollack, Daniel Ranard, Pratik Rath, Grant Remmen, Brandon Robinson, Simon Ross, and Brian Swingle for useful discussions, and we would like to thank Netta Engelhardt, Daniel Harlow, and Don Marolf for comments on the first draft of this letter.
N.B. is supported by the National Science Foundation under grant number 82248-13067-44-PHPXH, by the Department of Energy under grant number DE-SC0019380, and by New York State Urban Development Corporation Empire State Development contract no. AA289.
A.C.D. is a Postdoctoral Fellow (Fundamental Research) of the Research Foundation - Flanders, File Number 12ZL920N, and he was supported for part of this work by the KU Leuven C1 grant ZKD1118 C16/16/005, the National Science Foundation of Belgium (FWO) grant G.001.12 Odysseus, and by the European Research Council grant no. ERC-2013-CoG 616732 HoloQosmos.
B.E.N is supported by ERC grant ERC-2013-CoG 61673 HoloQosmos, and by the FWO and European Union's Horizon 2020 research and innovation program under the Marie Sk\l{}odowska-Curie grant agreement No. 665501. B.E.N. is an FWO [PEGASUS]$^2$ Marie Sk\l{}odowska-Curie Fellow. M.U. is supported by the National Science Foundation GRFP. This work was supported in part by the Berkeley Center for Theoretical Physics; by the Department of Energy, Office of Science, Office of High Energy Physics under QuantISED Award DE-SC0019380 and under contract DE-AC02-05CH11231; and by the National Science Foundation under grant PHY1820912. This material is based upon work supported by the National Science Foundation Graduate Research Fellowship Program under Grant No. DGE 1752814.

\bibliographystyle{utphys-modified}
\bibliography{refs}

\end{document}